\documentclass[11pt]{JHEP3}

\makeatletter
\@addtoreset{equation}{section}

\newcommand{\be}{\begin{equation}}
\newcommand{\ee}{\end{equation}}
\newcommand{\bear}{\begin{eqnarray}}
\newcommand{\eear}{\end{eqnarray}}
\newcommand{\ba}{\begin{array}}
\newcommand{\ea}{\end{array}}

\newcommand{\CA}{{\cal A}}

 \newcommand{\CV}{{\cal V}}

\preprint{ KIAS-P03092 \\ UOSTP-03106 \\ \tt hep-th/0312237}

\title{ Holography, Dimensional Reduction and the
Bekenstein Bound
}

\author{Dongsu Bak$^1$\footnote{\tt dsbak@mach.uos.ac.kr} and
Ho-Ung Yee$^2$\footnote{\tt ho-ung.yee@kias.re.kr}
\\
\\
$^1$Physics Department, University of Seoul, Seoul 130-743, Korea
\\
$^2$School of Physics, Korea Institute for Advanced Study, \\
{\it 207-43, Cheongryangri-Dong, Dongdaemun-Gu, Seoul 130-722,
Korea} }

\abstract{ \normalsize\noindent We consider dimensional reduction
of the lightlike holography of the covariant entropy bound from
$D+1$
dimensional  geometry of 
$M\times S^1$ to the D
dimensional geometry $M$. With a warping factor, the local
Bekenstein bound in $D+1$ dimensions leads to a more refined form
of the 
bound from the  D dimensional view point.
With this new local Bekenstein bound, it is quite possible to
saturate the lightlike holography 
even with nonvanishing expansion rate. With a Kaluza-Klein gauge
field, the dimensional reduction implies a stronger bound where
the energy momentum tensor contribution is replaced by the energy
momentum tensor with the electromagnetic contribution subtracted.
}

\begin{document}
\newpage

\section{Introduction}

The holographic principle appears to be a new guiding paradigm for
true understanding of quantum gravity theories. It states that the
fundamental  degrees of freedom in a certain region of spacetime
is bounded not by its volume but by its boundary area\cite{TH,SU}.
Based on the earlier attempts\cite{FS,BA}, a more  precise version
of the holography is proposed by Bousso\cite{BO}. In this
proposal, one considers the ``lightsheets'' consisting of
lightlike geodesic of nonpositive expansion orthogonally generated
from a boundary. The entropy passing through the lightsheet should
be bounded by the boundary area divided by the Planck area $4G$.
This statement is called the  covariant entropy bound
(CEB). This proposal passes many tests so far and is widely
accepted by now. Later, a more refined version was given by
Flanagan, Marolf and Wald\cite{FMW}. Here one considers the
lightlike sheets consisting of
 lightlike geodesics orthogonally generated
 from one boundary $B$ ending on another  boundary $B'$.
Then the proposal says
that the entropy passing through the lightlike sheet should be limited by
the difference of the boundary areas divided by $4G$. Hence this version is
certainly stronger than the original proposal by Bousso. This is the
formulation of the holographic principle which we shall mainly concern
in this note.

This refined  version may be
proved in a semiclassical context based on
 two conditions. One is the initial
condition on the lightlike sheet saying that
the minus of  initial expansion rate of the  geodesics divided by $4G$
should be taken
to be larger than the initial entropy flux density through
the lightsheet.
The other is the condition on the energy flux density  versus the change
of the entropy flux density. This condition is kind of the local version of
the Bekenstein bound of the entropy of weakly gravitating system\cite{BE},
\begin{equation}
S\  \le\  \pi M d
\end{equation}
where  $M$ and $d$ are respectively the energy and the linear size of
the system. From the local entropy energy condition,
 the Bekenstein bound may be derived straightforwardly. In this context,
we shall call the local entropy-energy condition
as ``local Bekenstein bound''.

As discussed in detail in \cite{Bak}, the first
 condition merely puts a
restriction on  the choice of the initial boundary surface.
This is because the initial expansion rate is solely determined
by the choice of the surface.
 Hence, at least classically, the condition that leads to the proof of the
statement comes from the local condition on the entropy and energy
densities.  The local Bekenstein bound is clearly a sufficient
condition for the holography. One unsatisfactory aspect about the
condition is that there is no way to saturate the holography bound
except some trivial case where the expansion remains zero all the
time\cite{Bak}. One typical example is the case of
AdS/CFT\cite{MA} where one compares the bulk gravitational entropy
with the regulated  boundary area divided $4G$ corresponding to the
degrees of freedom of the boundary CFT\cite{SusskindWitten}.
According to the above formulation of the holography, the
saturation cannot occur in the example of AdS/CFT
because there the expansion rate is nonvanishing for the
boundary surface we are interested in\cite{Bak}. This is
disappointing because the conjectured AdS/CFT correspondence
claims the exact equivalence at the level of Hilbert space. This
problem may disappear if the local Bekenstein bound is modified
appropriately, which we shall propose below.

In this note, we start from the CEB in $D+1$ dimensions and
investigate its implication upon dimensional reduction to the
spacetime of the form $M \times S^1$ where $S^1$ is the circle and
$M$ denotes a $D$ dimensional geometry\footnote{There are
literatures\cite{AA,BB} considering dimensional reduction of
holography in some other context. Ref.\cite{Hong:2003xd} discusses
phenomenological implications of entropy bounds in scenarios of
extra dimensions. See also Ref.~\cite{Karch} for 
an interesting application of the holography.}. 
We shall first consider the case where one
has vanishing Kaluza-Klein gauge fields. Lightlike geodesics
parallel to $M$ in the higher dimensions remain lightlike viewed
from lower dimensions. Assuming the local Bekenstein bound in
$D+1$ dimensions, the holography in the lower dimensions should
automatically follow because the dimensional reduction does not
change the statement of holography except some trivial overall
factor. What we are asking is then the implication of the higher
dimensional local Bekenstein bound. Considering
 the KK reduction with a warping factor, we find
that the $D+1$ dimensional local Bekenstein bound leads to a new
one with an order $G$ correction  from the view point of the lower
dimensional spacetime. This will be our main result. We further
consider the dimensional reduction with a KK gauge field but with
a trivial warp factor. This time we get a stronger version of
local Bekenstein bound, which seems consistent with previous
proposals in the presence of  $U(1)$ gauge field. The order $G$
correction of the Bekenstein bound from the dimensional reduction
should be considered seriously because the physics behind the
Bekenstein bound is not much understood now. Another virtue of
this correction is that the saturation of the bound is now
possible even with nonvanishing expansion.

In Section 2, we shall review the formulation of the lightlike
holography of CEB. In Section 3, we consider the dimensional
reduction with nontrivial warping factor or with a
KK 
gauge field.  We find interesting modifications of the local
Bekenstein bound. Last section is devoted for the discussions.

\section{Review of Covariant Entropy Bound}

In its original form proposed by Bousso\cite{BO},
the covariant entropy bound states that,
given a codimension two spacelike hypersurface, the total entropy on
a lightsheet generated by non expanding
null geodesics orthogonal to the hypersurface should be bounded
by the area of the surface in Planck unit. Null geodesics
may be either future
or past directed. Consider, for example, a future directed
focusing lightsheet,
which is orthogonal to a compact surface $\CA$ that
bounds a volume $\CV$.
The covariant entropy bound dictates the total entropy
flow on the lightsheet, which
is not less than the total entropy in the volume $\CV$ due
to the second law of
thermodynamics, be bounded by $\CA/4G$.
Thus, the covariant entropy bound generalizes the old area bound.
Another example would be taking a lightsheet generated by
past-directed null geodesics
that stay on the horizon of a growing black hole. The total
entropy flow on this lightsheet
is the total entropy of infalling matter while creating the black hole.
The covariant entropy bound dictates this be bounded by the area of
the horizon, which is
nothing but the Bekenstein-Hawking entropy of the black hole.
This makes a connection to the
generalized second law of thermodynamics (GSL).

In fact, trying to refine the connection to the GSL naturally
leads to a generalization of the original Bousso's
proposal\cite{FMW}. Imagine a situation where we have a black hole
and some matter is falling in it for a period of time. The GSL
implies that the black hole entropy increment (or equivalently,
its area increment) should not be less than the infalling matter
entropy. Consider the past-directed null geodesics staying on the
horizon as described in the above example, but now only during the
interval of matter infalling. The GSL in this case is equivalent
to saying that the total entropy flow on this converging
lightsheet must be bounded by the area difference between two
boundaries of the lightsheet. 
We may allow not only closed, but
also open codimension two hypersurfaces.

More precisely, let hypersurface orthogonal null geodesics
with affine parameter $\lambda$ be
generated by a null vector field $k^\mu$,
which is non expanding in the direction of $k^\mu$ i.e.
$\theta\equiv \nabla_\mu k^\mu \le 0$.  The vector
$k^\mu$ is either
future or past directed.
Denote the area of an orthogonal hypersurface
at an affine parameter $\lambda$ by $\CA(\lambda)$. Then, the
generalized covariant entropy bound
(GCEB) states that the total entropy,
$S(\lambda_f,\lambda_i)$,
on the lightsheet generated by
null geodesics with affine parameter range $[\lambda_i,\lambda_f]$,
should be bounded by $\Delta\CA/4G=(\CA(\lambda_i)-\CA(\lambda_f))/4G$,
\be
S(\lambda_f,\lambda_i)\leq \frac{1}{4G}
\Big(\CA(\lambda_i)-\CA(\lambda_f)\Big)\,.
\ee
The infinitesimal version\cite{Bak}
of this statement is
\be
s(\lambda)\leq-\frac{\theta(\lambda)}{4G}\,,
\ee
where $s(\lambda)\equiv - s_\mu k^\mu$ (future directed),
$+s_\mu k^\mu$ (past directed).

Bousso\cite{BO2002} showed that the GCEB actually
generalizes the Bekenstein bound, $S\leq \pi M d$.
There is a heuristic way to see how
the Bekenstein bound emerges from the GCEB.
Consider a weakly gravitating system with approximate
spherical symmetry of mass
$M$ and radius $R$, whose center is at the position zero of a
Cartesian system $\{x_{1,2,3}\}$.
Consider a  two-surface $\{x_2^2+x_3^2\leq R^2\}$ of
area $\pi R^2$ at $x_1=R$, and
imagine light rays orthogonal to and coming from this
surface toward the system.
While they are passing through the system, they are bent to converge
a little bit due to gravity effect. At the position $x_1=-R$, just behind the system,
their orthogonal  two surface would have a reduced radius,
\be
R'\approx R-\frac{1}{2}g(\Delta t)^2\,,
\ee
where $g=G M/R^2$ is the gravity acceleration at the radius $R$,
and $\Delta t =\frac{2R}{c}$
is the traveling time of the light rays. Hence
the contracted area at $x_1=-R$ is
\be
\CA'= \pi R'^2 \approx \pi\bigg(R-\frac{1}{2}\,\,
\frac{G M}{R^2}
\,\,
\Big(\frac{2R}{c}\Big)^2\bigg)^2
\approx \pi R^2-\frac{4\pi G M R}{c^2}\,,
\ee
where we assumed $R\gg\ 2G M/c^2$ as usual. Thus
\be
\frac{\Delta \CA}{4G}=\frac{\CA-\CA'}{4G}\approx \frac{\pi M R}{c^2}\,,
\ee
and this should bound the total entropy of the system, $S\leq\pi M R$ (in $c=1$ unit),
which is the Bekenstein bound.
The Bousso's analysis in Ref. \cite{BO2002}
refines the gravity effect using the Einstein's equation
for an arbitrary shape of the system. The result is a generalization of the Bekenstein
bound, $S\leq \pi M d$, where $d$ is now the smallest distance of any two parallel planes that
can enclose the system. This generalized Bekenstein bound may become very strong for thin systems.

In Refs. \cite{FMW,Bousso:2003kb,Strominger:2003br},
the authors suggested that the infinitesimal version
of the GCEB, $s(\lambda)\leq-\frac{\theta(\lambda)}{4G}$,
may be derived from certain assumptions,
\bear
i) \quad s(0) & \leq & -\frac{\theta(0)}{4G}\,,\\
ii)\quad \frac{ds}{d\lambda} &\leq & 2\pi T_{\mu\nu}k^\mu k^\nu\,,
\eear
where $T_{\mu\nu}$ is the energy-momentum tensor.
Note that the gradient assumption $ii)$  should hold for both
future- and past-directed null geodesics, i.e.
\be
\left|\frac{ds}{d\lambda}\right|
\ \leq\   2\pi T_{\mu\nu}k^\mu k^\nu
\ee
which implies that the weak energy
condition must hold. The derivation of the GCEB from $i)$ and $ii)$ is straightforward
using the Einstein's equation,
\be
R_{\mu\nu}-\frac{1}{2} g_{\mu\nu}\,R\,\,=\,\,8\pi G\, T_{\mu\nu}\,,
\ee
together with the Raychaudhuri's equation for null geodesics,
\be
\frac{d\theta}{d\lambda}\,=\,-\frac{\theta^2}{D-2}-\sigma_{\mu\nu}\sigma^{\mu\nu}+
\omega_{\mu\nu}\omega^{\mu\nu}-R_{\mu\nu}k^\mu k^\nu\,.
\ee
Note that for hypersurface orthogonal null geodesics, the twist $\omega_{\mu\nu}=0$ is absent.

The condition $ii)$ can be taken as an infinitesimal version of
the Bekenstein bound, leading to the usual Bekenstein bound when
integrated \cite{BO2002,Bousso:2003kb}, indicating that the bound
$ii)$ may be, in a sense, the most fundamental principle. To show
that $ii)$ leads to the usual Bekenstein bound, let us consider a
finite size, weakly gravitating system whose extension in $x_1$
direction is $\Delta x$. We call the other transverse dimensions
$\vec{y}$ collectively. We next imagine almost parallel light rays
(due to weak gravity) traveling along $x_1$ direction, and
normalize their affine parameter $\lambda$ such that \be \pm
k^\mu\,=\,\pm\frac{d\,x^\mu}{d\lambda_\pm}\,=
\,\bigg(\frac{\partial}{\partial t}+\frac{\partial}{\partial
x_1}\bigg)^\mu\,, \ee where $+k^\mu$ is for future-directed, while
$-k^\mu$ is for past-directed. Thus, $\lambda_+$ ($\lambda_-$) can
be identified to $x_1$ ($-x_1$). We let $\lambda_+$ run from
$x_1^i$ to $x_1^f$ and $\lambda_-$ from $x_1^f$ to $x_1^i$, where
$x_1^i$ ($x_1^f$) is the starting (ending) point of the system in
$x_1$ direction. Note that $ii)$ is independent of a normalization
of affine parameter. Because $s(x_1^{i,f})=0$, integrating $ii)$
for future directed case gives
\be
s(\lambda_+)\,\leq\,2\pi\int_{x_1^i}^{x_1} d\lambda_+
\,T_{\mu\nu}k^\mu k^\nu\,, \ee
while integrating for the
past-directed equation results in \be s(\lambda_-)\,\leq\,2\pi
\int_{\lambda_-(x_1^f)}^{\lambda_-(x_1)} d\lambda_-\,T_{\mu\nu}k^\mu k^\nu \,=\,2\pi
\int_{x_1}^{x_1^f} d\lambda_+ \,T_{\mu\nu} k^\mu k^\nu \,. \ee
Noting that $s(\lambda_+)=s(\lambda_-)=s(x_1)$, and summing the
above two equations lead to \be s(x_1)\,\leq\,\pi
\int_{x_1^i}^{x_1^f} dx_1 \,T_{\mu\nu} k^\mu k^\nu\,. \ee The
total entropy of the system is $S=\int d\vec y\int
dx_1\,s(x_1,\vec{y})$, hence we get \be S=\int
d{\vec{y}}\int_{x_1^i}^{x_1^f}dx_1\,s(x_1,\vec{y}) \leq\pi\Delta
x\int d\vec{y} \int dx_1\,T_{\mu\nu} k^\mu k^\nu=\pi\Delta x P_\mu
k^\mu\,, \label{derbe} \ee where $P_\mu=\int d\vec{y}\int dx_1
T_{\mu\nu} k^\nu$ is the total energy-momentum vector of the
system. For a static system, $P_0=M$, $P_i=0$, we obtain the
Bekenstein bound, $S\leq\pi \Delta x M$.

\section{Dimensional Reduction of Covariant Entropy Bound}

If the correct quantum theory of gravity is higher dimensional
and it indeed has the GCEB
as its fundamental aspect, we would have several
extra spatial dimensions.
One may expect that the GCEB in higher dimensions will
dimensionally reduce to the GCEB
in four dimensions for generic four dimensional observers
whose energy scale is
much smaller than the compactification scale.
However, some trace of the higher dimensional nature
of the fundamental theory
may manifest itself in certain modifications
of entropy bounds observed in lower dimensions.
In cases where the existence of an
additional dimension is only to provide
some specific lower dimensional situations, like a presence
of a KK 
$U(1)$ gauge field, the resulting modification of entropy bounds
in lower dimensions can be naturally attributed to a necessary
modification
of entropy bounds in lower dimensions in the
presence of these specific situations.
Namely  dimensional reduction plays a role of a
consistent tool to derive
these modifications.
In this section, we study the GCEB in $D+1$ spacetime dimensions,
with one spatial
dimension compactified on circle $S^1$ of  coordinate size $L$, and
describe the covariant entropy bound
that is relevant to $D$ dimensional observers.
By repeating this step, we may get down to any lower
dimensions from any higher dimensions.

We will consider two specific examples.
In the first case, we analyze the GCEB in $D+1$ dimensions formulated in terms of
two conditions $i)$ and $ii)$, in the space of an Einstein metric,
\be
ds^2\,=\, \tilde{g}_{MN}dx^M
dx^N\,=\,K(x
)^{\frac{1}{2-D}}
g^E_{\alpha\beta}\,\,
dx^\alpha dx^\beta
+K(x
)(dx^D)^2\,,
\ee
where $M,N=0,\ldots,D$ run over $D+1$
dimensions, $\alpha,\beta=0,\ldots,D-1$ are noncompact $D$ dimensional
indices, and $x^D$ is the compact $S^1$ direction.
The warp factor $K(x)$ as well as $g^E_{\mu\nu}$ is set to be
independent of $x^D$. With the above factorization, $ g_{\mu\nu}^E$
is the corresponding Einstein
metric in $D$ dimensions. Note that entropy bounds are formulated
in terms of Einstein metrics.
The second case we are going to study is the metric,
\be
ds^2\,=\, g^E_{\alpha\beta}\,\,
dx^\alpha
dx^\beta +(dx^D+ l\, A_\alpha dx^\alpha)^2\,,
\ee
where again everything is set to be independent of $x^D$, and $G$
is the Newton's constant in $D$ dimensions.
Here we introduce the length scale
$l$ defined
by $\sqrt{16\pi G}$. 
In $D$ dimensions, our system includes
a KK $U(1)$ gauge field  $ A_\alpha$. The above normalization of $A_\alpha$ produces the standard gauge kinetic term.
In the following,
we put tilde for every
relevant $D+1$ dimensional quantity.

We start from the two conditions, $i)$ and $ii)$, in $D+1$ dimensions,
\bear
&& i)\,\, \tilde{s}(0) \ \leq\  -\frac{\tilde\theta(0)}{4 \tilde{G}}\,, \\
&& ii)\,\,\frac{d \tilde{s}}{d\tilde\lambda}\  \leq\  2\pi \tilde{T}_{MN} {q}^M
{q}^N\,,
\eear
which lead to the $D+1$ dimensional GCEB,
\be
\tilde{s}(\tilde\lambda)\,\,\leq\,\,-
\frac{\tilde\theta(\tilde\lambda)}{4\tilde{G}
}\,,
\ee
where $q^M$ ($M=0,\ldots,D$)
are hypersurface orthogonal null geodesics,
$\tilde{s}=-s_M q^M$ is the entropy flow,
$\tilde\theta=\nabla_M q^M$
is the codimension two area contraction, $\tilde T_{MN}$ is
the energy-momentum tensor, and
$\tilde{G}$ is the Newton's constant, all in $D+1$ dimensions.
Note that $\tilde\lambda$ is the affine parameter with respect
to the $D+1$ dimensional
Einstein metric.
Now we take a codimension two hypersurface $\tilde{B}$
which is a direct
product of a codimension two hypersurface in $D$ dimensions
$B$ and $S^1$
in $x^D$ direction, i.e. $\tilde{B}=B\times S^1$.
Then, we choose hypersurface orthogonal $q^M$ of the form,
\be
q^M\,\,=\,\,(q^\alpha,\,0\,)\,
\ee
where $q^\alpha$ is independent of $x^D$ and is hypersurface
orthogonal to the $D$ dimensional
codimension two surface $B$.
For the metrics we are going to consider, it is easy to check that
$q^D=0$ is preserved, and moreover,
$q^\alpha$ is a null geodesic with respect to
the $D$ dimensional Einstein metric with an appropriate rescaling
of its affine parameter. Thus, we may formulate a $D$ dimensional
GCEB derived from the $D+1$ dimensional GCEB
for this rescaled $D$ dimensional
hypersurface orthogonal null geodesics. We will simply rewrite
the conditions $i)$ and $ii)$ in $ D+1$
dimensions in terms of correctly defined
$D$ dimensional quantities. These $D$ dimensional conditions,
$i')$ and $ii')$,
have modified expressions from the original proposal $i)$ and $ii)$.
Then, the mathematical procedure leading to
$\tilde{s}\leq -\tilde\theta/4\tilde{G}$
from $i)$ and  $ii)$ in $D+1$ dimensions guarantees that
we must have a corresponding mathematical proof of the $D$ dimensional
version of the GCEB from the modified conditions $i')$ and $ii')$ that are
obtained from dimensional reduction. Noting that the condition
$ii)$ (or $ii')$)
may be the most fundamental bound, the modifications
in $ii')$ we have obtained
in this paper should be considered seriously.

\subsection{The warped metric case}

We first consider the metric,
\be
ds^2=\tilde g_{MN}\,dx^M dx^N=K(x)^{\frac{1}{2-D}}
g^E_{\alpha\beta}\,dx^\alpha dx^\beta
+K(x)(dx^D)^2\,,
\ee
and a null geodesic vector field
\be
q^M\,=\,(q^\alpha(x),\,0\,)\,,
\ee
which is orthogonal
to $\tilde{B}=B \times S^1$, and independent of $x^D$.
From the null geodesic equation of $q^M$, it is easy to find that
\be
 k^\alpha\,\equiv\,K^{\frac{1}{2-D}}\, q^\alpha\,,
\ee
is a correctly normalized $D$
dimensional null geodesic in terms of $g^E_{\mu\nu}$.
In other words, the affine parameter $\lambda$ with respect
to $ g^E_{\mu\nu}$ is given by
\be
\frac{d}{d \lambda}\,=\,K^{\frac{1}{2-D}}\,\frac{d}{d\tilde\lambda}\,,
\ee
where $\tilde\lambda$ is the affine parameter of $q^M$.
Thus,
\bear
\tilde\theta&\equiv&\nabla_M q^M\,=\,\partial_\alpha
q^\alpha+\tilde\Gamma^M_{M\alpha} q^\alpha\nonumber \\
&=&\partial_\alpha\Big(K^\frac{1}{D-2} k^\alpha\Big)+
\tilde\Gamma^M_{M\alpha}
\Big(K^\frac{1}{D-2}  k^\alpha\Big)\,,
\eear
where $\tilde\Gamma$ is the affine connection of $g_{MN}$.
Now, writing $\tilde\Gamma^M_{M\alpha}$ in terms of
$ g^E_{\mu\nu}$,
\bear
\tilde\Gamma^D_{D \alpha}&=&\frac{1}{2}\partial_\alpha \log K\nonumber\\
\tilde\Gamma^\beta_{\beta\alpha}&=&
\Gamma^\beta_{\beta\alpha}+\frac{D}{2(2-D)}\,\partial_\alpha \log K\,,
\eear
where $\Gamma^\alpha_{\beta\gamma}$ is the affine connection of
$g^E_{\mu\nu}$, we have
\be
\tilde\theta\,=\,
K^\frac{1}{D-2}\,
(\partial_\alpha  k^\alpha+
\Gamma^\beta_{\beta\alpha}  k^\alpha)
\,=\,K^\frac{1}{D-2} \theta\,,
\ee
where $\theta=\nabla_\alpha  k^\alpha$ is the area
contraction of $ k^\alpha$ in terms of
$g^E_{\mu\nu}$. Also,
\bear
\tilde{s}&=& -\tilde s_M\,q^M\,= -\,\tilde{s}_\alpha\,K^\frac{1}{D-2}
k^\alpha\nonumber\\
&=&-\frac{ s_\alpha
k^\alpha}{L}\,K^\frac{1}{D-2}\,=\,\frac{ s}{L}\,K^\frac{1}{D-2}\,,
\eear
where $s_\alpha\equiv L \, \tilde{s}_\alpha$ is
the $D$ dimensional entropy current density.
Using these relations, we can simply rewrite the condition $i)$ and the
GCEB in $D+1$ dimensions in terms of $D$  dimensional quantities,
\be
\tilde{s}(\tilde\lambda)\,\leq\,-\frac{\tilde\theta(\tilde\lambda)}
{4 \tilde{G}}\ \longleftrightarrow\
 s( \lambda)\,\leq\,-\frac{\theta(\lambda)}{4G}\,,
\ee
where $L\, G=\tilde G$ is the Newton's constant in $D$
dimensions. Henceforth, the condition $i)$
and the GCEB will trivially reduce to $D$ dimensions.

However, a difference arises in dimensional reduction of
the condition $ii)$. Recall that $\tilde T_{MN}$ satisfies the $D+1$
dimensional Einstein equation,
\be
\tilde R_{MN}\,-\,\frac{1}{2}\tilde
 g_{MN}\,\tilde R\,=\,8\pi \tilde G \,\tilde T_{MN}\,,
\ee
while the $D$
dimensional energy-momentum tensor $T_{\alpha\beta}$
is defined to satisfy the Einstein's
equation with the metric $g^E_{\mu\nu}$,
\be
R_{\alpha\beta}\,-\,\frac{1}{2} g^E_{\alpha\beta}\, R\,=
\,8\pi G\,T_{\alpha\beta}\,.
\ee
Because $q^M$ ($ k^\alpha$) is null, we have
\bear
8\pi \tilde G\,\tilde T_{MN} q^M q^N &=& \tilde R_{MN}\,q^M q^N\,,
\nonumber\\
8\pi G\, T_{\alpha\beta} k^\alpha
 k^\beta &=& R_{\alpha\beta}\, k^\alpha  k^\beta\,.
\eear
Now  writing $\tilde R_{\alpha\beta}$
in terms of the metric $g^E_{\mu\nu}$,
\be
\tilde R_{\alpha\beta}\,=\,
R_{\alpha\beta}-\frac{\ \,(D-1)}{4(D-2)}\,(\partial_\alpha\log K)
(\partial_\beta \log K)
+\frac{1}{2(D-2)}\Big(\nabla_\gamma \nabla^\gamma\log K\Big)\,
g_{\alpha\beta}^E\,,
\ee
and $q^\alpha=K^\frac{1}{D-2}
k^\alpha$, we get
\be
8\pi \tilde G\,\tilde T_{MN}q^M q^N\,=\,K^\frac{2}{D-2}\bigg\{
8\pi G\, T_{\alpha\beta} k^\alpha  k^\beta\,
-\,\frac{(D-1)}{4(D-2)}\,\Big(\frac{d\log K}{d \lambda}\Big)^2\bigg\}\,.
\ee
Also,
\be
\frac{d \tilde s} {d\tilde\lambda}\,=\,\frac{K^\frac{1}{D-2}}{L}
\frac{d}{d  \lambda}\Big(
 s\, K^\frac{1}{D-2}\Big)\,=\,\frac{K^\frac{2}{D-2}}{L}
\bigg\{\frac{d  s}{d\lambda}+\frac{s}{D-2}
\frac{d \log K}
{d\lambda}
\bigg\}\,.
\ee
Using these relations, the condition $ii)$, $d \tilde s/
d\tilde\lambda\leq 2\pi
\tilde T_{MN} q^M q^N$, is equivalent to
\be
ii')\,\frac{d
s}{d\lambda}\,
\leq\,2\pi T_{\alpha\beta} k^\alpha  k^\beta
-\frac{\phantom{abc}(D-1)}{16 G(D-2)}\bigg(\frac{d\log K}{d \lambda}\bigg)^2
-\frac{ s}{(D-2)}\bigg(\frac{d\log K}{d \lambda}\bigg)\,.
\ee
Note that once we extract $ T_{\mu\nu}$
from the relevant Einstein equation, $K$ can be an arbitrary
function of $x^\alpha$, which is invisible to $D$
dimensional observers. In fact,
the most interesting case is when we get the weakest condition
from $ii')$ by suitably choosing $K$.
This is achieved by completing square of the correction
terms in the rhs of $ii')$,
\be
\Delta=
-\frac{1}{16 G}
\frac{(D-1)}{(D-2)}\bigg(\frac{d\log K}
{d\lambda}+\frac{8 G\, s } {(D-1)}\bigg)^2
+\frac{4G}{(D-1)(D-2)} s^2\,.
\ee
In other words,
\be
ii'')\  \frac{d  s}{d\lambda}\ \leq\
2\pi
 T_{\alpha\beta} k^\alpha k^\beta
+\frac{4G}{(D-1)(D-2)}\, s^2
\ee
is the weakest condition we
derived from dimensional reduction.
We stress that the GCEB in $D+1$ dimensions
guarantees that $i)$ and  $ii'')$ should
imply the GCEB in $D$ dimensions. Imagining dimensional reduction
cascade from infinite dimension to the $D$ dimensions, we can
actually strengthen $ii'')$ to
\be
ii''')\, \frac{d s}{d \lambda}\,\leq\,
2\pi  T_{\alpha\beta} k^\alpha  k^\beta
+\frac{4G}{(D-2)}\, s^2\,.
\ee

In fact, using the Einstein equation and the
Raychaudhuri equation, we can prove this in a purely
$D$ dimensional point of view. The point is that the correction term
in $ii''')$
is taken care of by the $\frac{\theta^2}{D-2}$ term in the
Raychaudhuri equation, which was previously ignored.
Explicitly, using  condition $ii''')$, the Einstein equation
 and $s(\lambda) \le
-\theta(\lambda)/(4G)$, one has
\be
{ds\over d\lambda}\ \le\  2\pi T_{\alpha\beta} k^\alpha k^\beta
+{4G\over D-2} s^2 \ \le\
 {1\over 4G} \left( R_{\alpha\beta} k^\alpha k^\beta +{\theta^2\over D-2}
\right) \ \le\  -{\phantom{iii}d\theta\over 4G\, d\lambda}
\ee
where the Raychaudhuri equation is used for the last
inequality. Combining with the initial condition,  the holography
$s(\lambda) \ \le\
-\theta(\lambda)/4G$ is proved self
consistently in purely D dimensional
point of view.
One thing to note is
that we assumed $s\ge 0$ for the proof here.

\subsection{The case of a KK $U(1)$ gauge field}

We next consider the metric
\be
ds^2\,=\,  g^E_{\alpha\beta}
dx^\alpha dx^\beta +(dx^D+ l\, A_\alpha dx^\alpha)^2\,,
\ee
which gives a KK $U(1)$ gauge field
$A_\alpha$ upon dimensional reduction.
The corresponding modification
of the entropy bound in $D$
dimensions can be considered as a necessary modification
of the entropy bound in the presence of a $U(1)$ gauge interaction.

The null geodesic equations of this metric written in terms of the $D$
dimensional Einstein metric $ g^E_{\mu\nu}$ are
( $q^M=(k^\alpha,q^D)$ )
\bear
&&  \frac{d k^\alpha}{d\lambda}+\Gamma^\alpha_{\beta\gamma}
 k^\beta  k^\gamma =
l\,
Q\,F^\alpha\,_\beta\, k^\beta\,,\nonumber\\
&& q^D+ l 
\,A_\alpha k^\alpha =  Q\,=\,{\rm constant}\,,
\eear
with the constraint
\be
 g^E_{\alpha\beta}k^\alpha k^\beta +Q^2\,=\,0\,.
\ee
The $F_{\alpha\beta}=\partial_\alpha A_\beta-\partial_\beta A_\alpha$ is the
usual field strength.
Indices are raised or lowered by $ g^E_{\mu\nu}$ always.
For null geodesics with $Q=0$, we see that $k^\alpha$ is
a null geodesic in $D$ dimensions with respect to $ g^E_{\mu\nu}$.
Moreover, we can always choose the gauge such that
\be
A_\alpha k^\alpha\,=\,0\,.
\ee
In this gauge, $q^D=0$ and we have
$q^M=(k^\alpha,0)$ as before. This enables us to take $\tilde B= B\times S^1$
as our codimension two hypersurface and $k^\alpha$ is hypersurface
orthogonal to $B$. Hence we may
formulate the $D$ dimensional GCEB for $k^\alpha$.

The $D+1$
dimensional area contraction
is $\tilde\theta=\nabla_M q^M=
\partial_\alpha k^\alpha+\tilde\Gamma^M_{M\alpha}k^\alpha$,
and writing $D+1$ dimensional connection $\tilde\Gamma$
in terms of $D$ dimensional quantities, it is straightforward to
get $\tilde\Gamma^M_{M\alpha}=\Gamma ^\beta_{\beta\alpha}$, which gives
$\tilde\theta=\nabla_\alpha k^\alpha= \theta$.
We also have
\be
\tilde s\,=\,- \tilde s_M q^M\,=- \,\tilde s_\alpha k^\alpha\,=\,-
\frac{ s_\alpha k^\alpha}{L}\,=\,
\frac{ s}{L}\,,
\ee
where $s$ is the entropy flux in $D$ dimensions.
Using these relations,
\be
\tilde s(\tilde\lambda)\,\leq\,-\frac{\tilde\theta(\tilde\lambda)}
{4 \tilde G}\ \longleftrightarrow\   s(\lambda)\,\leq\,
-\frac{\theta(\lambda)}{4G}\,,
\ee
where $L\,  G=\tilde G$. Thus, again the condition $i)$
and the GCEB are the same as usual.

An interesting modification arises in the condition $ii)$, however.
Working out the Ricci tensor in $D+1$
dimensions in terms of $D$ dimensional quantities,
\be
 \tilde R_{\alpha\beta}\,=\, R_{\alpha\beta}+
{l^2\over 2}
\Big(
-F_{\gamma\alpha}F^\gamma\,_\beta
+A_\alpha\nabla_\gamma F_\beta\,^\gamma+A_\beta\nabla_\gamma
F_\alpha\,^\gamma\Big)+\frac{ 
l^4}
{4}A_\alpha A_\beta \, F_{\eta\gamma}F^{\eta\gamma} \,,
\ee
and remembering the gauge $A_\alpha k^\alpha=0$, we have
\bear
8\pi \tilde G \, \tilde T_{MN} q^M q^N &=& \tilde R_{MN}q^M q^N\,=\,
R_{\alpha\beta}k^\alpha k^\beta
-8\pi G F_{\gamma\alpha}k^\alpha F^\gamma\,_\beta k^\beta \nonumber\\
&=& 8\pi G \Big( T_{\alpha\beta} k^\alpha k^\beta - F_{\gamma\alpha}
k^\alpha F^\gamma\,_\beta k^\beta\Big) \,.
\eear
This gives us
\be
2\pi \tilde T_{MN} q^M q^N =\frac{1}{L}\bigg\{2\pi
T_{\alpha\beta}k^\alpha k^\beta
-2\pi
F_{\gamma\alpha}k^\alpha\,\, F^\gamma\,_\beta k^\beta \bigg\}\,,
\label{ema}
\ee
and the condition $ii)$ is dimensionally reduce to
\be
ii')\, \frac{d s}{d\lambda}\,\leq\,2\pi\left(
 T_{\alpha\beta}
-F_{\gamma\alpha}\,\,F^\gamma\,_\beta
\right)k^\alpha k^\beta
\,.
\ee
We propose that this is the correct version of the
Bekenstein bound in the presence of
a $U(1)$ gauge interaction. This kind of correction is not
new. In the presence of electromagnetic charge,
there appeared a proposal\cite{EM},
\be
S \le 2 \pi R \left( E- {Q^2\over 2 R}\right)
\label{emb}
\ee
where $R$ is the radius of the spherically symmetric system and
Q is the charge. The modification appearing in
our bound has the  same structure because one is
subtracting the electromagnetic contribution from the total
energy-momentum tensor. 
It is clear that our proposal corresponds to the local
version of (\ref{emb}).



The correction term is negative definite. To show this, let us work
in a local Lorentzian frame where the metric becomes flat with $g_{\mu\nu}
={\rm diag} (-1,1,\cdots,1)$. It follows that,
\be
F_{\gamma\alpha}k^\alpha\,\,F^\gamma\,_\beta k^\beta=
F_{0i} F_{0i} k_j k_j - (F_{0i} k_i)^2 + F_{ij}k_j F_{ik} k_k \ge 0\,,
\ee
where we used the fact $k_0^2= k_i k_i$.
Hence this part of the correction gives a tighter bound than the original
proposal.

\section{Physical Implication of the refined local Bekenstein bound}

In the previous section, we obtained the refined version of
local Bekenstein bound in D dimensions
\be
{ds\over d\lambda} \  \le \  2\pi \, T_{\mu\nu} k^\mu k^\nu
+ {4G\over D-2} \, s^2
\label{localbound}
\ee
through dimensional reduction. Once we accept the conditions
$i)$ and $ii)$
in arbitrary dimensions for the lightlike holography, the above
follows naturally as we have seen in the previous section.

First of all, the correction is of next order in the Newton's constant
$G$. Note that the validity of the original Bekenstein bound was
argued mainly for weakly gravitating systems. Since our
modification is of order $G$, this new bound is quite consistent
with previous investigations in this respect.

The second point is that now the condition allows a nontrivial saturation
of the holography bound. Consider, for example,
 a system with spherical symmetry,
 which implies that the shear, $\sigma_{\mu\nu}$, is
vanishing. It is clear that the lightlike holography can be
saturated if the new local entropy bound is
saturated and if $s(0)= -\theta (0)/(4G)$. The condition of zero expansion
is no longer required.

To illustrate the above point, let us
consider the  AdS$_{5}\times S^5$ geometry dual to the N=4 SYM
theory. The AdS metric in the Poincare patch reads
\be
ds^2= R_{AdS}^2\left({-dt^2+ dz^2+ d\vec{y}^2 \over z^2}
\right)
\ee
where 
$\vec{y}$ is a $D-2$ dimensional coordinate with $D=5$ for the AdS$_5$.
According to the AdS/CFT correspondence, the entropy of  AdS bulk  counted by the
boundary area divided by $4 G_5$ ought to saturate the
maximal
entropy of the boundary CFT\cite{SusskindWitten}.
Thus, for the maximal  capacity of bulk entropy current,
the saturation
of the holography bound is necessary for the AdS/CFT
correspondence to hold.
The boundary of interest
is located
at  $z=\delta$, which is related  to
the UV cut-off energy scale of the SYM theory
by  $\Lambda=1/\delta$
via the UV/IR connection. The geodesic orthogonally
generated from the boundary is described by
\be
k^\mu= {z^2\phantom{ii}\over R^2_{AdS}}\,\,(1,\ 1,\ \vec{0})
\ee
with $k^\mu= (k^0, k^z,\vec{k}^y)$. Along the geodesic,
the affine parameter $\lambda$ is related to the coordinate by
\be
z= -{R^2_{AdS}\over \lambda \phantom{ii}}
\label{affine}
\ee
with a range $(-{R^2_{AdS}/\delta},\  -0)$ for
$z \in (\delta, \infty)$. In the transverse space of $\vec{y}$, one has
\be
\nabla_n k_m = {1\over z} \,\,\delta_{nm}
\ee
where $n$ and $m$ are indices for the $\vec{y}$ directions.
Hence $\sigma_{nm}=\omega_{nm}=0$
and $\theta= - (D-2) \,{z\,/ R^2_{AdS}}$. Therefore the saturation
of the holography bound implies
\be
s=-{\theta\over 4G_5}=  {(D-2) \over 4 G_5}\,\,
{z\phantom{ii}\over R^2_{AdS} }=
- {(D-2) \over 4 G_5}\,\,{1\over  \lambda}
\ee
Since the system is time independent,
the entropy current density is then
\be
s^\mu = {(D-2) \over 4 G_5}\,\,{z\phantom{ii}\over R^2_{AdS} }
\,\, (1,\ 0,\ \vec{0})\,,
\ee
which is consistent with the scale symmetry of the AdS
geometry.
One may check that the above
entropy  density saturates our local bound
(\ref{localbound}) as
\be
{ds\over d\lambda}={4 G_5\over D-2}\,\, s^2 =
{(D-2) \over 4 G_5}\,\,{1\over \lambda^2} \,,
\ee
with
$T^{AdS}_{\mu\nu} k^\mu k^\nu=0$ that follows from the fact
$T^{AdS}_{\mu\nu} \sim g_{\mu\nu}$.
For $\lambda \in (
\lambda_i, \lambda_f)$, the
integrated form of the entropy flux through the lightsheet
becomes
\be
S(\lambda_f,\lambda_i)\equiv \int d \vec{y}
\int^{\lambda_f}_{\lambda_i} d \lambda\,\, \sqrt{h}
 \,\, s \,\,
e^{\int^\lambda_{\lambda_i} d\lambda' \theta(\lambda')}
={1\over 4 G_5} \int d\vec{y}\,\, 
\,\,{|\lambda_i|^{D-2}- |\lambda_f|^{D-2}\over R^{D-2}_{AdS} } \,,
\ee
which agrees with $({\cal A}_i- {\cal A}_f)/4G_5 $. Here $h$ is the
determinant of the induced metric for the boundary surface.
With the choice of   $\lambda_f=-0$,
the lightsheet covers the whole coordinate patch and
$S= {\cal A}_i/4G_5$, which is the desired relation for the holography.
Thus, the bulk entropy surrounded
by the $z=\delta$ boundary can be consistently saturated by the Bekenstein-Hawking entropy
formula. There is a nonsupersymmetric dilatonic variation of
 AdS$_5\times S^5$, which is called the ``Janusian'' background\cite{SH}.
It would be interesting to ask if saturation is again possible
for this background.

We now ask the implication of the refined local version to the
Bekenstein bound.  For simplicity, let us consider the case
when $s$ is
uniform. Then by a similar computation leading to (\ref{derbe}),
one finds
\be
S \ \le\  \pi d E + {4G\over D-2} {d^2\over 2} A\, s^2 =\pi d E  +
{4G\over 2(D-2)}\,\,{S^2 \over A} \,.
\ee
The condition can be further solved in terms of $S$ by
\be
S \ \le\  \pi d E {2\over 1+\sqrt{1 - {8G \over D-2}{\pi d E\over A}} }\ = \
\pi d E  +
{4G\over 2(D-2)}\,\,{(\pi d E)^2 \over A}+\cdots  \,\,.
\ee
Clearly the correction is of order $G/A$, which comes up to our expectation.
However we do not understand its physical origin.
Further clarification of the refined local
Bekenstein bound  is necessary.

%

\section{Discussions}
In this note, we considered dimensional reduction of the
lightlike holography from $D+1$ geometry of the form
$M\times S^1$. With a warping factor, the local Bekenstein bound
in $D+1$ dimensions leads to a more refined form of the local
Bekenstein bound from the view point of the D dimensional
geometry. With  a KK gauge field, dimensional
reduction leads to the stronger bound where  the energy
momentum tensor contribution is replaced by the energy momentum
tensor with the electromagnetic contribution subtracted.
This local version is consistent with the previously
proposed modification of the Bekenstein bound in the
presence of a non-vanishing electromagnetic  charge.

With the order $G$ correction of the refined local
Bekenstein bound,
we argued that saturation of the holography bound
is now quite possible even for nonvanishing
expansions.  For the AdS/CFT discussed in the
introduction, we showed that the regulated boundary
area divided
$4G$ agrees with the degrees
of freedom of the boundary CFT.

There has been some challenge in finding examples violating
the Bekenstein bound. Our proposal suggests that  one way to find
a violation of the original version is to look for possible corrections
of order $G$. Further understanding of the refined version of the
local Bekenstein bound would be very interesting.

\vskip 2cm
\centerline{\large \bf Acknowledgment}
  We are grateful to Seok Kim
for the contributions at the earlier stage of this work.
D.B. is supported in part by
Korea Research Foundation Grant KRF-2003-070-C00011.
H.-U.Y. is supported in part by KOSEF
R01-2003-000-10319-0.

\end{document}